Jean Bricmont
FYMA, 2, ch. du Cyclotron
B-1348 Louvain-la-neuve, Belgium
Tel:32-10-473277
Fax: 32-10-472414

\documentstyle[12pt]{article}

\catcode `\@=11
\@addtoreset{equation}{section}
 \catcode `\@=12
\newcommand{\be}{\begin{eqnarray}}
\newcommand{\bea}{\begin{eqnarray*}}
\newcommand{\en}{\end{eqnarray}}
\newcommand{\ena}{\end{eqnarray*}}
\newcommand{\non}{\nonumber}
\newcommand{\no}{\noindent}
\newcommand{\vs}{\vspace}
\newtheorem{Th}{Theorem}

\newtheorem{Pro}{Proposition}

\title{{\bf Global Large Time Self-similarity of a  Thermal-Diffusive
Combustion System with Critical Nonlinearity}}

\author{J.Bricmont\thanks{Supported by EC grant CHRX-CT93-0411} \\ UCL,
Physique Th\'eorique, B-1348, Louvain-la-Neuve, Belgium
\\bricmont@fyma.ucl.ac.be\and A.Kupiainen\thanks {Supported by NSF grant
DMS-9205296  and EC grant CHRX-CT93-0411} \\ Helsinki University,
Department of Mathematics,\\ Helsinki 00014, Finland
\\ajkupiai@cc.helsinki.fi \and J. Xin\thanks{Supported by
 NSF grant DMS-9302830 and Swedish NFR Grant F-GF10448-301} \\Department of
Mathematics, University of Arizona,\\ Tucson,
 AZ 85721, USA\\xin@math.arizona.edu}

\date{}

\begin{document}

\maketitle \begin{abstract}

We study the initial value problem of the thermal-diffusive combustion
system: $u_{1,t} = u_{1,x,x} - u_1 u^2_2, u_{2,t} = d u_{2,xx} + u_1 u^2_2,
x \in R^1$, for non-negative spatially decaying initial data of arbitrary
size and for any positive constant $d$. We show that if the initial data
decays to zero sufficiently fast at infinity, then the solution $(u_1,u_2)$
converges to a self-similar solution of the reduced system: $u_{1,t} =
u_{1,xx} - u_1 u^2_2, u_{2,t} = d u_{2,xx}$, in the large time limit. In
particular, $u_1$ decays to zero like ${\cal O}(t^{-\frac{1}{2}-\delta})$,
where $\delta > 0$ is an  anomalous exponent depending on the initial data,
and $u_2$ decays to zero with normal rate ${\cal O}(t^{-\frac{1}{2}})$. The
idea of the proof is to combine the a priori estimates for the decay of
global solutions with the renormalization group (RG) method for
establishing
the self-similarity of the solutions in the large time limit.
\end{abstract}

\section{Introduction.}

\vs{5mm}

In this paper, we study the initial value problem of
 the following thermal-diffusive combustion system: \be u_{1,t} &=&
u_{1,xx}
- u_1 u^2_2 \label{a1}\\ u_{2,t} &=& du_{2,xx} + u_1 u^2_2, \hspace*{10mm}
x
\in R^1, \label{a2} \en with non-negative initial data $(u_1,u_2)|_{t=0} =
(a_1(x),a_2(x)) \in (L^1 (R^1)\cap L^\infty (R^1))^2$, of arbitrary size,
 where the positive constant $d$ is the Lewis number. We are interested in
the large time behavior of solutions of (\ref{a1})-(\ref{a2}).

The system (\ref{a1})-(\ref{a2}) on a bounded domain is
 well-studied in the literature, see \cite{1}, \cite{8}, \cite{9},
\cite{10}
and references therein. In case of homogeneous Dirichlet or Neumann
boundary
conditions, the large time behavior of solutions is that $(u_1,u_2)$
converges uniformly to a constant vector $(c_1,c_2)$ such that $c_1 \cdot
c_2 =0$, see K. Masuda \cite{10}.

More recently, the system (\ref{a1})-(\ref{a2}) on the line $R^1$ has been
proposed and investigated as a model for cubic autocatalytic chemical
reactions of the type $A + 2B \to 3B$, with rate proportional to
$u_1u^2_2$,
where $u_1$ and $u_2$ are concentrations of the reactant $A$, and the
autocatalyst $B$. We refer to the interesting papers by J. Billingham and
D.
Needham \cite{4}, \cite{5}, for details. In \cite{4} and \cite{5}, the
authors established the existence of traveling front solutions rigorously
by
shooting and phase plane methods; moreover, they studied the long time
asymptotics of solutions by formal methods and numerical computations for a
class of front initial data, i.e. data such that $a_1 + a_2$ has
nonvanishing limits as $x \to \infty$.

Motivated by thermal-diffusive models with Arrhenius reactions, \cite{11},
\cite{2} etc. Berlyand and Xin \cite{3} considered system
(\ref{a1})-(\ref{a2}) for a class of small initial data in $(L^1 \cap
L^\infty (R^1))^2$ and showed that $u_i (i=1,2)$ are bounded from above and
below by self-similar upper and lower solutions. The results of \cite{3}
imply that $u_1$ decays to zero in time with an algebraic rate faster than
$t^{-\frac{1}{2}-\delta}$, for some $\delta > 0$, and $u_2$ decays to zero
like ${\cal O}(t^{-\frac{1}{2}})$.

In the present work, we prove the exact large time self-similar asymptotics
with no restriction on the size of initial data as long as the data has
sufficiently fast spatial decay. Our main result is the following. We
consider the system (\ref{a1})-(\ref{a2}) with initial data $(a_1,a_2) \in
{\cal B} \times {\cal B}$, where ${\cal B}$ is the Banach space of
continuous functions on ${R}^1$ with the norm \be \| f \| = \sup_{x\in R^1}
| f(x)| (1+|x|)^q , \;\; \mbox{with} \; q > 1 \;\; \mbox{fixed below}.
\label{a3} \en Let $\phi = \phi(x)$ be the Gaussian: \be \phi (x) =
\frac{1}{\sqrt{4\pi d}} e^{-\frac{x^2}{4d}}. \label{a4} \en Given $A \geq
0$, let $\psi_A$ be the principal eigenfunction (ground state) of the
differential operator: \be {\cal L}_A = - \frac{d^2}{dx^2} - \frac{1}{2} x
\frac{d}{dx} - \frac{1}{2} + A^2 \phi^2 (x), \label{a5} \en on $L^2 (R^1,
d\mu)$, with $d\mu(x)=e^{\frac{x^2}{4}} dx$. The corresponding eigenvalue
is
denoted by $E_A \geq 0$ (and $E_A > 0$ for $A>0$). We normalize $\psi_A$ by
$\int \psi_A^2 (x) d\mu (x) =1$. Our main result is:

\vspace*{5mm} \par\noindent \begin{Th} (Global Large Time Self-Similarity).
Consider initial data $(a_1,a_2) \in {\cal B} \times {\cal B}, a_i \not
\equiv 0, a_i \geq 0, i=1,2$. Let $A = \int_{R^1} a_1(x) + a_2 (x) dx$, the
total mass of the system, which is conserved in time. Then system
(\ref{a1})-(\ref{a2}) has a unique global classical solution
$(u_1(x,t),u_2(x,t)) \in {\cal B} \times {\cal B}$ for $\forall t \geq 0$.
Moreover, there exists a $q(A)$ such that, if $q \geq q(A)$ in (\ref{a3}),
there is positive number $B$ depending continuously on $(a_1,a_2)$ such
that: \end{Th} \be && \| t^{\frac{1}{2} + E_A} u_1 (\sqrt{t}\cdot, t) - B
\psi_A (\cdot) \|_{\stackrel{\longrightarrow}{t \uparrow \infty}}0,
\label{a6}\\ && \| t^{\frac{1}{2}} u_2 (\sqrt{t}\cdot, t) - A \phi (\cdot)
\|_{\stackrel{\longrightarrow}{t \uparrow \infty}}0. \label{a7} \en

\vspace*{5mm} \par\noindent {\bf Remark 1.1} All the results of Sections 2
and 3 hold for any $q>1$. We need the decay at infinity of $a_i$ to be fast
enough only to obtain the exact decay rate in (\ref{a6}). For $A$ large,
$E_A$ will be large, and the decay in (\ref{a6}) may be much faster than
the
diffusive one. Alternatively, we could consider data $a_i \in {\cal
B}_{\exp}$ where ${\cal B}_{\exp}$ is defined through the norm $$ \| f
\|_{\exp} = \sup_x | f(x)| e^{\gamma|x|} $$ for some $\gamma >0$. Then, the
conclusions of Theorem 1 hold for any $A$.

\vspace*{5mm} \par\noindent {\bf Remark 1.2} The rate of convergence in
(\ref{a6}) and (\ref{a7}) to zero is actually $O(t^{-\eta})$,  for some
$\eta>0$, see (\ref{d27}, \ref{d28}). The convergence in (\ref{a6}) and
(\ref{a7}) implies that $$ u_1(x,t) \sim \frac{B \psi_A (\frac{x}{\sqrt
t})}{t^{\frac{1}{2}+E_A}} + h.o.t. \;\;\;\;\; \mbox{and} $$ $$ u_2(x,t)
\sim
\frac{A}{t^{\frac{1}{2}}} \phi (\frac{x}{\sqrt t}) + h.o.t., $$ as $t \to
\infty$, where the leading terms are just the two parameter self-similar
solutions to the reduced system. The anomalous exponent $E_A$ occurs as a
result of the interactions of nonlinearities of opposite signs.
Furthermore,
$E_A$ can be computed or estimated as the ground state energy of operator
 ${\cal L}_A$ depending only on the Lewis number $d$ and the total mass of
the system. A nonperturbative upper bound is  $\frac{A^2}{4 \pi
\sqrt{2d+1}}$, while, for $A$ small, $$ E_A = \frac{A^2}{4\pi \sqrt{2d+1}}
-
O (A^4), $$ see \cite{3} for details. Actually, it is more natural
physically to normalize $A=1$, which amounts to putting a coupling constant
$A^2$ in front of the reaction terms in (\ref{a1},\ref{a2}), and the
anomalous exponent $E_A$ depends then on the strength of that coupling
constant.

\vspace*{5mm} \par\noindent {\bf Remark 1.3} In order to understand the
heuristics of (\ref{a6}), (\ref{a7}), consider a more general problem: \be
u_{1,t} &=& u_{1,xx} - u_1 u^m_2 \label{a8} \\ u_{2,t} &=& du_{2,xx} + u_1
u^m_2 \label{a9} \en for $m \geq 1$. For $m>2$, as explained in \cite{3},
we can use the RG method of \cite{6} to prove that both $u_1$ and $u_2$ go
diffusively to zero. For $1 \leq m < 2$, one can use the maximum principle,
as in Lemma 2.3 and equation (\ref{c5}) below, to bound from above $u_1$ by
$\bar u_1$, which is the solution of \be \bar u_{1,t} = \bar u_{1,xx} -
{\cal O} (t^{-m/2} \phi^m (\frac{x}{\sqrt{t}})) \bar u_1 \label{a10} \en
Then using the Feynman-Kac formula, we get \be u_1 (x,t) \leq \exp (-{\cal
O} (t^{1-\frac{m}{2}})) \label{a11} \en
 for $|x| \leq {\cal O} (\sqrt t)$. For $|x| \geq {\cal O} (\sqrt t)$, one
gets a diffusive behaviour, depending on the rate of decay, as $x \to \pm
\infty$, of the initial data. Then, inserting the fast decay (\ref{a11}) of
$u_1$ in (\ref{a9}), one shows that the effect of the nonlinear term in
(\ref{a9}) is small and that $u_2$ diffuses to zero. Clearly the borderline
case $m=2$ is the most delicate and the most interesting one. Instead of
(\ref{a11}), one gets $\exp (-{\cal O} (\log t))$ which gives rise, after
some analysis, to (\ref{a6}).

\vspace*{5mm} The rest of the paper is organized as follows.  In section 2,
we derive a priori estimates on the solutions of the system
(\ref{a1})-(\ref{a2}) based on the work of K. Masuda \cite{10} for finite
domains. Quite a few estimates are different here due to the unboundedness
of $R^1$. The a priori estimates imply the existence of global smooth
solutions. In section 3, we derive decay estimates for the solutions using
the maximum principle and a simple renormalization group (RG, see \cite{6})
idea to show that $u_1$ goes to zero like ${\cal
O}(t^{-\frac{1}{2}-\delta})$, for some $\delta > 0$. We use this
information
to prove that the nonlinearity is irrelevant (in the RG sense) in
(\ref{a2})
and that $\| u_2 \|_\infty \leq O(t^{-\frac{1}{2}})$ as $t \to \infty$. In
section 4, we use the results of sections 2 and 3, and the renormalization
group method to prove the convergence to a self-similar solution and thus
complete the proof of the main theorem.

\vs{5mm}

\section{A priori estimates and global bounds.}

\vs{5mm}

The goal of this section is to prove \vspace*{3mm} \begin{Pro} The system
(\ref{a1}, \ref{a2}) has a unique classical solution satisfying \be \| u_i
\|_{L^p} \leq C (a_1, a_2) ,\;\; i=1,2, \;\; 1 \leq p \leq + \infty,
\label{b1}\en where the constant C depends only on the initial data
$(a_1,a_2) \in (L^1 (R^1)\cap L^\infty (R^1))^2$. \end{Pro}

\par\noindent {\bf Remark 2.1} Although some of the arguments below follow
those of Masuda \cite{10}, we provide them for completeness. Here and
below,
we use $C$ to denote a generic constant that may vary from place to place.
Moreover we write, as above, $C(\cdot)$ to indicate the only variables on
which the constant may depend.

\vspace*{5mm} First, we have the obvious

\vspace*{3mm} \par\noindent {\bf Lemma 2.1} The solution $(u_1,u_2)$
satisfies the $L^1$ estimates:  \be \| u_1 + u_2 \|_{L^1(R^1)} = \| a_1 +
a_2 \|_{L^1(R^1)}&,& \;\;\; \| u_2 \|_{L^1(R^1)} \geq \| a_2 \|_{L^1
(R^1)},
\non \\ \| u_1 \|_{L^1(R^1)} \leq \| a_1 \|_{L^1(R^1)}&,& \;\;\;
\int^\infty_0 \int_{R^1} u_1 u^2_2 dx d \tau < + \infty.\label{b2} \en

\vs{5mm}\par\noindent {\bf Proof} $\;\;$ Integrating (\ref{a1})-(\ref{a2})
over $R^1$, assuming spatial decay at infinity, we get: \be \| u_1 \|_{L^1}
(t) = \| a_1 \|_{L^1} - \int^t_0 \| u_1 u^2_2 \|_{L^1} (\tau) d \tau,
\label{b3}\\ \| u_2 \|_{L^1} (t) = \| a_2 \|_{L^1} + \int^t_0 \| u_1 u^2_2
\|_{L^1} (\tau) d \tau. \label{b4} \en Combining (\ref{b3})-(\ref{b4})
gives
(\ref{b2}). \hfill$
\makebox[0mm]{\raisebox{0.5mm}[0mm][0mm]{\hspace*{5.6mm}$\sqcap$}}$ $
\sqcup$

\vs{5mm}\par\noindent {\bf Lemma 2.2} The function $g_p(u_2) \equiv u^p_2$
satisfies, for $p\geq 2$, \be 0 \leq g'_p (u_2) \leq ((\frac{p}{p-1}) g_p
(u_2) g''_p (u_2))^{\frac{1}{2}}. \label{b5} \en

\vs{5mm}\par\noindent {\bf Proof} $\;\;$ Direct calculation. \hfill$
\makebox[0mm]{\raisebox{0.5mm}[0mm][0mm]{\hspace*{5.6mm}$\sqcap$}}$ $
\sqcup$

\vs{5mm} Using the classical parabolic maximum principle, we have,

\vs{3mm}\par\noindent{\bf Lemma 2.3} \be &(1)& 0<u_1(x,t) \leq \| a_1
\|_\infty, \;\; \forall \; t>0; \label{b6}\\ &(2)& 0<\underline{u}_2 (x,t)
\leq u_2(x,t), \;\; \forall \; t>0, \;\; \mbox{where} \; \underline{u}_2
\;\;
\mbox{is a solution of:} \nonumber\\ && \underline{u}_{2,t} = d
\underline{u}_{2,xx}, \;\; \underline{u}_2|_{t=0} = a_2(x); \label{b7}\\
&(3)& u_1(x,t) \leq \bar{u_1} (x,t) , \;\; \forall t \geq 0, \;\;
\mbox{where}\; \bar{u_1} \;\; \mbox{solves:} \nonumber\\ &&\bar{u_1}_t =
\bar u_{1,xx} - \bar{u_1} \cdot \underline{u}^2_2, \;\; \bar{u_1}|_{t=0} =
a_1(x).\label{b8} \en

\vs{5mm}\par\noindent {\bf Remark 2.2} Using this Lemma, one immediately
proves (\ref{b1}) for $u_1$, since, by (\ref{b8}), $u_1$ is a fortiori
bounded by the solution of the heat equation with the same initial data.

\vs{5mm}\par\noindent{\bf Lemma 2.4} The solutions $(u_1,u_2)$ of
(\ref{a1})-(\ref{a2}) satisfy the $L^p$ bounds:  \be \| u_i \|_{L^p} \leq
C(a_1,a_2,p) < + \infty, \;\; i=1,2, \;\;1 \leq p < + \infty, \;\;
\mbox{and} \; p = \; \mbox{integer}. \label{b9} \en

where $C(a_1,a_2,p)$ is a constant depending only on the initial data and
$p$.

\vs{5mm}\par\noindent {\bf Proof} Due to Remark 2.2, we have only to prove
the bounds for $u_2$. We use standard local existence of classical
solutions
for parabolic equations, and, therefore, we freely integrate by parts
below.
Our goal will be to  prove bounds uniform in time. We shall show that

\be &&\int_{R^1} u^p_2 dx + \int^t_0 \int_{R^1} (u_{2,x}^2 u_2^{p-2}
+u_{1,x}^2 u_2^{p} + u_1u_2^{p+2})dx d \tau \nonumber\\ &\leq& C(a_1,a_2,p)
(1+\int^t_0 \int_{R^1}u_1u_2^{p+1}dx d \tau) \label{b10} \en for all $p\geq
2$ ( $p$ integer). Besides, we shall show, for $p=1$, \be \int^t_0
\int_{R^1} ( u_1u_2^{3}+u_{1,x}^2 u_2 )dx d \tau &\leq& C(a_1,a_2)
(1+\int^t_0 \int_{R^1}u_1u_2^{2}dx d \tau) \label{b11} \en Using (\ref{b2})
to bound $\int^t_0 \int_{R^1}u_1u_2^{2}dx d \tau$, and using induction in
$p$, we get that all the terms on the left hand side of (\ref{b10}) are
bounded, for all $p\geq 1$, $p$ integer ($u_2 \in L^1$ by (\ref{b2})). In
particular, this implies the claims of the Lemma.

To prove (\ref{b10}, \ref{b11}), we multiply (\ref{a2}) by $g'_p (u_2)$, we
integrate over $R^+_1 \times R_1$, and we get, after integrating by parts:
\be \int_{R^1} g_p (u_2) dx &=& \int_{R^1} g_p (a_2) dx - d \int^t_0
\int_{R^1} u_{2,x}^2 g''_p (u_2) dx d \tau \nonumber\\ &+& \int^t_0
\int_{R^1} u_1u^2_2 g'_p (u_2) dx d\tau,\label{b12} \en here $g_p(u_2) =
u^p_2, \;\;p \geq 1$. \\ Consider the identity $(p\geq 1)$, \be &&
\frac{d}{dt} \int_{R^1} (u^2_1 + u_1) g_p (u_2) dx = \int_{R^1}
(1+2u_1)(u_{1,xx}-u_1u^2_2)g_p(u_2)dx \nonumber \\ && + \int_{R^1}
(u_1+u^2_1) g'_p (u_2) (du_{2,xx} + u_1u^2_2)dx \nonumber\\ &=& -
\int_{R^1}
(1+2u_1) u_{1,x} u_{2,x} g'_p (u_2) dx - 2 \int_{R^1} u^2_{1,x} g_p (u_2)
dx
- d \int_{R^1} (1+2u_1)u_{1,x}u_{2,x}g'_p(u_2)dx \nonumber\\ &-& d
\int_{R^1}
(u_1+u^2_1) g''_p (u_2) u^2_{2,x} dx - \int_{R^1} (1+2u_1)
u_1u^2_2g_p(u_2)dx+\int_{R_1}(u_1+u^2_1)g'_p(u_2)u_1u^2_2dx \nonumber \\
&\equiv& I+II+III+IV+V+VI.\label{b13} \en We estimate for $p \geq 2$, using
(\ref{b6}), (\ref{b5}): \be &&I+II+III \leq
(1+2\|a_1\|_\infty)(1+d)\int_{R^1}|u_{1,x}u_{2,x}| g'_p (u_2) dx - 2
\int_{R^1} u^2_{1,x} g_p (u_2) dx \nonumber\\ &\leq& (1+2 \|a_1\|_\infty)
(1+d) \int_{R^1} | u_{1,x} u_{2,x}| (g_p (u_2) g''_p
(u_2)(\frac{p}{p-1}))^{\frac{1}{2}} dx \nonumber \\ && - 2 \int_{R^1}
u_{1,x}^2 g_p (u_2) dx \nonumber\\ &\leq& \frac{1}{2} (1+2
\|a_1\|_\infty)(1+d)( \frac{p}{p-1})^{\frac{1}{2}} (\epsilon \int_{R^1}
u_{1,x}^2 g_p (u_2) dx + \epsilon^{-1} \int_{R^1} u_{2,x}^2 g''_p(u_2) dx)
\nonumber\\ && - 2 \int_{R^1} u_{1,x}^2 g_p (u_2) dx\label{b14} \en Picking
$$ \epsilon = \frac{2}{(1+2\|a_1\|_\infty)(1+d)}
(\frac{p-1}{p})^{\frac{1}{2}} $$ in (\ref{b14}), we continue: \be &&
I+II+III \nonumber\\ &\leq & ( \frac{p}{p-1} ) \epsilon^{-1}
(1+2||a_1\|_\infty)^2 (1+d)^2 \int_{R^1} u_{2,x}^2 g''_p (u_2) dx -
\int_{R^1} u_{1,x}^2 g_p (u_2) dx \label{b15} \en

In addition, we have: \be IV &\leq & 0 , \label{b16} \\ V &\leq& -
\int_{R^1} u_1 u^2_2 g_p (u_2) dx, \label{b17} \\ VI &\leq& (\|a_1\|_\infty
+ \|a_1\|^2_\infty) \int_{R^1} g'_p (u_2) u_1 u^2_2 dx. \label{b18} \en
Integrating (\ref{b13}) from zero to $t$ yields: \be &&0\leq ( \int_{R^1}
(u_1+u^2_1)g_p(u_2)dx) (t) \leq \int_{R^1} (a_1 + a^2_1) g_p (a_2) dx
\non\\
&& + C(p,a_1) \int^t_0 \int_{R^1} u^2_{2,x} g''_p (u_2) dx d\tau
\nonumber\\
&+& (\| a_1 \|_\infty + \|a_1\|^2_\infty) \int^t_0 \int_{R^1} g'_p (u_2)
u_1
u^2_2 dx d\tau \nonumber\\ &-& \int^t_0 \int_{R^1} u_1 u^2_2 g_p (u_2) dx d
\tau -\int_{R^1} u_{1,x}^2 g_p (u_2) dx d\tau . \label{b19} \en Combining
(\ref{b12}) and (\ref{b19}) gives (\ref{b10}) for $p\geq 2$.

For $p=1$, we proceed from (\ref{b13}) as follows:  \be && \frac{d}{dt}
\int_{R^1} (u_1+u^2_1) u_2 dx = - \int_{R^1} (1+2u_1)u_{1,x}u_{2,x}dx-2
\int_{R^1} u^2_{1,x} u_2 dx \nonumber \\ &&-d \int_{R^1} (1+2u_1) u_{1,x}
u_{2,x} dx - \int_{R^1} (1+2u_1) u_1 u^3_2 dx \nonumber \\ &&+ \int_{R^1}
(u_1 + u^2_1) u_1u^2_2 dx \nonumber \\ &\leq& (1+2 \| a_1\|_\infty)
\int_{R^1} |u_{1,x} u_{2,x}| dx - 2 \int_{R^1} u^2_{1,x}u_2 dx \nonumber \\
&&+ d (1+2\|a_1\|_\infty) \int_{R^1} |u_{1,x} u_{2,x}| dx - \int_{R^1} u_1
u^3_2 dx \nonumber \\ &&+ (\|a_1\|_\infty + \|a_1 \|^2_\infty) \int_{R^1}
u_1 u^2_2 dx \nonumber \\ &\leq& \frac{1}{2} (1+d) (1+2 \| a_1 \|_\infty)
[\epsilon^{-1} \int_{R^1} u_{1,x}^2 dx + \epsilon \int_{R^1} u^2_{2,x} dx ]
\nonumber \\ && - 2 \int_{R^1} u^2_{1,x}u_2 dx - \int_{R^1} u_1 u^3_2 dx
\nonumber \\ && + (\| a_1\|_\infty + \| a_1\|^2_\infty) \int_{R^1} u_1
u^2_2
dx, \label{b20} \en for  any $\epsilon > 0$. Now, integrate (\ref{b20})
from
$0$ to $t$, to get (for $\epsilon $ small enough): \be && \int^t_0
\int_{R^1} (2u_1 u^3_2 + u_{1,x}^2 u_2) dxd\tau \leq 2d \int^t_0 \int_{R^1}
u_{2,x}^2 dxd\tau \nonumber\\ &+& C(a_1,a_2) (1+ \int^t_0 \int_{R^1}
(u_{1,x}^2 + u_1 u_2^2) dxd\tau). \label{b21} \en Now, use (\ref{b12}) with
$p=2$ to bound \be 2d \int^t_0 \int_{R^1} u_{2,x}^2 dxd\tau\leq \int_{R^1}
a^2_2 dx + \int^t_0 \int_{R^1} u_1 u_2^3 dxd\tau. \label{b22} \en Finally,
observe that, multiplying (\ref{a1}) by $u_1$, and integrating by parts, we
get: \be \frac{d}{dt} \int_{R^1} u_{1}^2 dx =-\int_{R^1} u_{1,x}^2 dx-
\int_{R^1} u_1^2 u_2^2 dx \label{b23} \en from which we immediately obtain:
\be \int^t_0 \int_{R^1} u_{1,x}^2 dx d\tau\leq \int_{R^1} a^2_1 dx
\label{b24} \en uniformly in $t$. Combining (\ref{b21}, \ref{b22},
\ref{b24}), we get (\ref{b11}).  This completes the proof of the lemma.
\hfill$ \makebox[0mm]{\raisebox{0.5mm}[0mm][0mm]{\hspace*{5.6mm}$\sqcap$}}$
$
\sqcup$

\vspace*{5mm} \par\noindent {\bf Remark 2.3} In Masuda \cite{10},
$g_p(u_2)=(1+u_2)^p$, and lemma 2.4 is proved for fractional $p$'s by
starting the induction from (\ref{b12}), $p\in (0,1)$ instead of combining
(\ref{b20}) and (\ref{b19}). However, for unbounded domains like $R^1$,
such
argument fails since $g_p(u_2) \not \in L^1(R^1)$. Lemma 2.4 still holds
for
fractional powers $p$, and we will show that with the help of $L^\infty$
bounds which we establish below using (\ref{b10}).

\vs{5mm}\par\noindent {\bf Lemma 2.5} The solutions $(u_1,u_2)$ of
(\ref{a1})-(\ref{a2}) obey the estimates \be \|u_{1,x}\|_2 + \|u_{2,x}\|_2
\leq { C} (a_1,a_2,a_{1,x},a_{2,x}) \label{b25} \en where $C
(a_1,a_2,a_{1,x},a_{2,x})$ is a positive constant depending on
$\|a_i\|_1,\|a_i\|_\infty, \|a_{i,x}\|_2$.

\vs{5mm}\par\noindent {\bf Proof} Multiplying (\ref{a1}) by $u_{1,xx}$,
(\ref{a2}) by $u_{2,xx}$, and integrating over $R^1$ gives: $$ -
\frac{1}{2}
\frac{d}{dt} \|u_{1,x}\|^2_2 = \int_{R^1} |u_{1,xx}|^2 dx - \int_{R^1} u_1
u^2_2 u_{1,xx} dx $$ or \be \frac{1}{2} \frac{d}{dt} \| u_{1,x} \|^2_2 = -
\int_{R^1} |u_{1,xx}|^2 dx - \int_{R^1} u^2_2 u^2_{1,x} dx - 2 \int_{R^1}
u_2u_1 u_{1,x} u_{2,x} dx \label{b26} \en Similarly, we have: \be
\frac{1}{2} \frac{d}{dt} \| u_{2,x} \|^2_2 = - d \int_{R^1} |u_{2,xx}|^2 dx
+ 2 \int_{R^1} u_1 u_2 u^2_{2,x} + \int u^2_2 u_{1,x} u_{2,x} dx.
\label{b27} \en Adding (\ref{b26}) and (\ref{b27}) gives: \be \frac{1}{2}
\frac{d}{dt} (\| u_{1,x} \|^2_2 + \| u_{2,x}\|^2_2 ) = - \int_{R^1}
u_{1,xx}^2 dx - d \int_{R^1} u^2_{2,xx} dx - \int_{R^1} u^2_2 u^2_{1,x} dx
\non\\ - 2 \int u_1 u_2 u_{1,x} u_{2,x} dx + 2 \int u_1u_2u^2_{2,x} dx +
\int u^2_2 u_{1,x}u_{2,x}dx. \label{b28} \en Now, integrate (\ref{b28})
from
$0$ to $t$, and  bound the resulting terms on the right hand side as
follows. The first three terms are negative, and the last three terms can
be
bounded, using $\|u_1\|_\infty \leq \| a_1 \|_\infty$ (see (\ref{b6})) and
the Cauchy-Schwarz' inequality: \be | \int_0^t \int_{R^1} u_1u_2 u_{1,x}
u_{2,x} dx d\tau | \leq \| a_1 \|_\infty (\int_0^t \int_{R^1}
u^2_{1,x}u^2_2
dx d\tau \int_0^t \int_{R^1} u^2_{2,x} dx d\tau) ^{1/2} \label{b29} \en \be
| \int_0^t \int_{R^1} u_1u_2 u^2_{2,x} dx d\tau | \leq \| a_1 \|_\infty
\int_0^t \int_{R^1} u_2 u^2_{2,x} dx d\tau, \label{b30} \en and \be |
\int_0^t \int_{R^1} u^2_2 u_{1,x} u_{2,x} dx d\tau | \leq (\int_0^t
\int_{R^1} u^2_2 u^2_{1,x} dx d\tau\int_0^t \int_{R^1} u^2_2 u^2_{2,x} dx
d\tau)^{1/2}. \label{b31} \en Now, the terms on the right hand side of
(\ref{b29}, \ref{b30}, \ref{b31}) are uniformly bounded in $t$, because all
the terms in the left hand side of (\ref{b10}) are bounded, for $p \geq 2$.
So, we get \be \| u_{1,x} \|^2_2 (t) + \| u_{2,x} \|^2_2 (t) \leq {C}
(a_1,a_2,a_{1,x},a_{2,x}), \forall \; t \geq 0.  \label{b32} \en \hfill$
\makebox[0mm]{\raisebox{0.5mm}[0mm][0mm]{\hspace*{5.6mm}$\sqcap$}}$ $
\sqcup$

\vspace*{5mm} \par\noindent By local existence of classical solutions in $C
(\left[0,T^\ast\right.);(L^2(R^1))^2) \cap { C} ((0,T^\ast)$;
$(H^2(R^1))^2)$ of system (\ref{a1})-(\ref{a2}), and parabolic regularity:
$\exists t_1 >0$, such that $$ \| u_{1,x}\|^2_2 + \| u_{2,x}\|^2_2 (t_1)
\leq C (t_1)( \|a_1\|^2_2 + \| a_2\|^2_2), \;\; t_1 < T^\ast $$ If we
replace 0 by $t_1$ in the proof of Lemma 2.5, then (\ref{b25}) in fact
implies

\vspace*{5mm}\par\noindent {\bf Corollary 2.1} \be \| u_{1,x} \|_2 + \|
u_{2,x} \|_2 \leq { C} (a_1,a_2) \label{b33} \en {\it where the constant
$C$
depends only on $\|a_i\|_1$ and} $\|a_i\|_\infty, i=1,2$.

\vspace*{5mm} \par\noindent {\bf Corollary 2.2} {\it By Sobolev imbedding,
(\ref{b9}) for $p=2$ and (\ref{b33}) imply: \be \|u_2\|_\infty \leq { C}
(a_1,a_2). \label{b34} \en } Thus Proposition 1 follows from:

\vspace*{5mm} \par\noindent {\bf Corollary 2.3} {\it Combining the
estimates
in Lemma 2.4, (\ref{b6}), (\ref{b34}) and standard local existence of
classical solutions, we have shown that global smooth and bounded solutions
exist for system (\ref{a1})-(\ref{a2}) in $(L^1 (R^1) \cap L^\infty
(R^1))^2$.}

\vs{5mm}

\section{Decay estimates}

\vs{5mm}

To exhibit the decay of the solutions of (\ref{a1})-(\ref{a2}),  let us
introduce the scaled solutions $$ \tilde u_1 (x,t) = \sqrt{t} u_i (\sqrt t
x, t) $$ for $i=1,2$. From now on, we consider nonnegative initial data
$(a_1(x),a_2(x)) \in {\cal B} \times {\cal B} \subseteq (L^1 (R^1)\cap
L^\infty (R^1))^2$ for $q>1$. The purpose of this section is to prove

\vspace*{5mm} \begin{Pro} The solution $(u_1,u_2)$ of (\ref{a1})
constructed
in Proposition 1 satisfies the bounds \be \| \tilde u_1 \| (t) &\leq& { C}
(\|a_1\|,\|a_2\|) (1+t)^{-\delta} \label{c1} \\ \| \tilde u_2\| (t) &\leq&
C
(\|a_1\|,\|a_2\|) \label{c2} \en where $\delta = \delta(\|a_2 \|_{L^1}) >
0$
and $\|\cdot\|$ is, for all $t>0$, the norm (\ref{a3}). \end{Pro}

\vspace*{5mm} \par\noindent {\bf Remark 3.1} Note that, in particular,
(\ref{c1}) and (\ref{c2}) imply  \be \|u_1\|_\infty (t) \leq C
(\|a_1\|,\|a_2\|) (1+t)^{-\frac{1}{2}-\delta},\nonumber\\
 \|u_2\|_\infty (t) \leq C (\|a_1\|,\|a_2\|) (1+t)^{-\frac{1}{2}}
\label{c3}
\en

\vspace*{5mm} \par\noindent {\bf Remark 3.2} The bound (\ref{c1}) was
essentially derived in \cite{3}, using (\ref{b8}) and (\ref{b7}), but with a
different norm.

\vs{5mm}\par\noindent {\bf Proof} $\;$ Using (\ref{b8}), (\ref{b7}), it is
enough to prove (\ref{c1}) with $u_1$ replaced by $\bar{u_1}$. Also, since
$\underline{u_2}$ solves the heat equation, we have: \be \underline{u_2}
(x,t) = \frac{A}{\sqrt t} \phi ( \frac{x}{\sqrt t} ) + h.o.t. \label{c4}
\en
as $t \to \infty$, with $\phi$ given by (\ref{a4}) and $A = \int_{R^1} a_2
(x) dx$.

The higher order terms are easy to control, so we can consider, instead of
(\ref{b8}), \be \bar{u}_{1,t} = \bar u_{1,xx} - \frac{A^2}{t} \phi^2 (
\frac{x}{\sqrt t}) \bar u_1 \label{c5} \en By a simple change of variables,
$\xi = \frac{x}{\sqrt t}, \tau = \log t$, we get, for $t \geq 1$, \be \sqrt
t \bar u_1 (x,t) = (e^{-\tau {\cal L}_A} \bar u_1 (\cdot,1)) (\xi)
\label{c6} \en with ${\cal L}_A$ given by (\ref{a5}). Then, (\ref{c1})
follows from the first bound  on the semigroup $e^{-\tau {\cal L}_A}$ given
in Lemma 4.1.b.\\  To prove (\ref{c2}), we need

\vs{5mm}\par\noindent {\bf Lemma 3.1}{\it (Sharp Decay of $u_2$ in
$L^\infty$ norm). There exists a constant $C$ depending only on $\|a_i\|,\;
i=1,2$ such that } \be \|u_2\|_\infty (t) \leq
\frac{C}{(1+t)^{\frac{1}{2}}}, \;\; \forall t \geq 0. \label{c7} \en
Consider then equation (\ref{a2}): $u_{2,t} = d u_{2,xx} + (u_1u_2) u_2$.
By
the first inequality in (\ref{c3}) (which follows from (\ref{c1})), and
Lemma 3.1, $u_2 \leq \bar{u_2}$, where $\bar{u_2}$ solves: \be \bar u_{2,t}
&=& d \bar u_{2,xx} + C (1+t)^{-1-\delta} \bar u_2 \non\\ \bar u_2 |_{t=0}
&=& a_2. \label{c8} \en But (\ref{c8}) is a linear heat equation, from
which
(\ref{c2}) follows easily since $\int^\infty_0 (1+t)^{-1-\delta} dt <
\infty$. \hfill$
\makebox[0mm]{\raisebox{0.5mm}[0mm][0mm]{\hspace*{5.6mm}$\sqcap$}}$ $
\sqcup$

\vs{5mm} We are left with the

\vs{5mm}\par\noindent{\bf Proof of Lemma 3.1} Write equation (\ref{a2}) in
integral form: \be u_2 (t,x) &=& \int_{R^1} \frac{1}{\sqrt{4\pi dt}} e^{-
\frac{x^2}{4dt}} a_2 (x-y) dy + \non \\ & & \int^t_0 \int_{R^1} (4\pi
ds)^{-\frac{1}{2}} \exp ( - \frac{y^2}{4ds}) (u_1u^2_2) (x-y,t-s) dy ds.
\label{c9} \en Taking $L^\infty$ norm in (\ref{c9}) yields \be \| u_2
\|_\infty (t) \leq \frac{C (\|a_2\|)}{(1+t)^{\frac{1}{2}}} + C \int^t_0
s^{-\frac{1}{2}} \|u_1u^2_2\|_{L^1} (t-s)ds \label{c10} \en

Now, use $\|u_1u^2_2\|_{L^1} \leq \|u_1\|_\infty \|u^2_2\|_{L^1}, \| u_1
\|_\infty (t) \leq C (\|a_1\|,\|a_2\|) (1+t)^{-1/2-\delta}$ (which is the
first inequality in (\ref{c3})), \be \| u^2_2 \|_{L^1} (t) \leq \|u_2
\|_\infty (t) \| u_2 \|_{L^1} (t) \leq C (\|a_1\|,\|a_2\|), \label{c11} \en
(which follows from Proposition 1) and \be \int^t_0 s^{-\frac{1}{2}}
(1+(t-s))^{-1/2-\delta} ds \leq C(1+t)^{-\delta} \label{c12} \en to get \be
\| u_2 \|_\infty (t) \leq C (\|a_1\|,\|a_2\|) (1+t)^{-\delta} \label{c13}
\en (for $\delta \leq 1/2$, or else, (\ref{c7}) is already proven). Now,
use
(\ref{c13}) to improve (\ref{c11}) into $\| u^2_2 \|_{L^1} (t) \leq C
(\|a_1\|,\|a_2\|) (1+t)^{-\delta}$, which, inserted into (\ref{c10}) yields
$$ \| u_2 \|_\infty (t) \leq C(\|a_1\|,\|a_2\|) (1+t)^{-2\delta}, $$ and we
can iterate up to $1/2$, which proves (\ref{c7}). \hfill$
\makebox[0mm]{\raisebox{0.5mm}[0mm][0mm]{\hspace*{5.6mm}$\sqcap$}}$ $
\sqcup$

\vs{5mm}

\section{Self-similarity}

\vs{5mm}

In this section we apply the Renormalization Group  method to improve
Proposition 2 and finish the proof of the Theorem. We prove:

\begin{Pro} Under the assumptions of Theorem 1, there exists $\epsilon > 0$
such that, if \be \| a_1 \| \| a_2\| < \epsilon, \label{d1} \en the claims
of Theorem 1 hold. \end{Pro}

\vs{5mm}\par\noindent {\bf Proof of Theorem 1} By Proposition 2, we can
find
a $T<\infty$ such that the functions $u_{iT} (x,t) = \sqrt T u_i (\sqrt T
x,
Tt)$ satisfy \be \| u_{1T} (\cdot, 1) \|\| u_{2T} (\cdot, 1) \| < \epsilon,
\label{d2} \en where $T$ depends on the initial data $(a_1,a_2)$. Moreover,
$u_{iT}$ solve the equations (\ref{a1}) and (\ref{a2}), and thus, by
Proposition 3 and (\ref{d2}), $u_{iT}$ and hence $u_i$, will have the
asymptotics claimed in the Theorem. \hfill$
\makebox[0mm]{\raisebox{0.5mm}[0mm][0mm]{\hspace*{5.6mm}$\sqcap$}}$ $
\sqcup$

\vs{5mm}

We will now set up an inductive scheme for the proof of Proposition 3.  We
define, for $L>1$, \be u^{(n)}_i (x,t) = L^n u_i (L^n x, L^{2n} t),\;\; t
\in [1,L^2]. \label{d3} \en Then $u^{(n)}_i$ satisfy the equations
(\ref{a1}) with initial data \be a^{(n)}_i (x) = L^n u_i (L^nx,L^{2n}).
\label{d4} \en We will study $a^{(n)}_i$ inductively in $n$ i.e. we will
consider the RG map $(a_1,a_2) \to (a'_1, a'_2)$ defined in ${\cal B}
\times
{\cal B}$ where \be a'_i (x) = Lu_i (Lx,L^2) \label{d5} \en and $(u_1,u_2)$
solve (\ref{a1})-(\ref{a2}) with initial data $(a_1,a_2)$. We first prove a
Lemma for the linearization of this map when (\ref{a1})-(\ref{a2}) is
linearized around the expected asymptotics (\ref{a6}) and (\ref{a7}).

Hence we consider the equations \be v_{1t} &=& v_{1xx} - A^2 t^{-1} \phi^2
(
\frac{x}{\sqrt t}) v_1 \non\\ v_{2t} &=& dv_{2xx} \label{d6} \en for $t\in
[1,L^2]$ and $v_i(x,1) = a_i (x)$. By the change of variables $\xi =
\frac{x}{\sqrt t}, \tau = \log t$ one gets \be Lv_1 (Lx,L^2) &=&
(L^{-{2\cal
L}_A} a_1) (x) \non \\ L v_2 (Lx,L^2) &=& (L^{-{2\cal L}} a_2) (x)
\label{d7} \en where \be {\cal L} = - d \frac{d^2}{dx^2} - \frac{1}{2} x
\frac{d}{dx} - \frac{1}{2} \label{d8} \en and ${\cal L}_A$ is given by
(\ref{a5}). Recall also that $\psi_A$ is the principal eigenvalue (ground
state) of ${\cal L}_A$. We collect some properties of ${\cal L}_A$ and
${\cal L}$ in

\vs{5mm}\par\noindent {\bf Lemma 4.1} {\it ${\cal L}_A$ and ${\cal L}$ have
the following properties } \begin{enumerate}{\it \item[a)] ${\cal L}_A
\psi_A = E_A \psi_A, \;\; E_A >0\;\;$ if $A>0$. \item[b)] Let $f \in {\cal
B}$.
 There exist $\delta > 0$ and $\tau_0<\infty$ such that, for $\tau \geq
\tau_0$, $$ \| e^{-\tau{\cal L}_A} f \| \leq e^{-\tau\delta} \|f\|. $$
Moreover, there exists $q(A)$ such that,  if $f\in {\cal B}$, with
$(\psi_A,f) = 0$ ($(\cdot,\cdot)$ being the scalar product in ${\cal H}=L^2
({\bf R}, d\mu)$, $d\mu=e^{\frac{x^2}{4}}dx$), and with $q>q(A)$ in
(\ref{a3}), $$ \| e^{-\tau{ \cal L}_A} f \| \leq e^{-\tau(E_A+\delta)}
\|f\|
$$ \item[c)] Let $P_A$ be the orthogonal projection in ${\cal H}$ on
$\psi_A$. The quantities $|E_A - E_{A'}|, |1-(\psi_A,\psi_{A'})|, \| P_A
-P_{A'}\|$ (operator norm in ${\cal B}$) are bounded by $C(K)|A-A'|$ and
$\|P_A\| \leq C(K)$, for $0\leq A,A' \leq K$. \item[d)] $e^{-\tau {\cal L}}
\phi = \phi$. \item[e)] Let $f \in {\cal B}, \int f dx = 0$. Then, there
$\delta > 0$ and $\tau_0<\infty$ such that, for $\tau \geq \tau_0$,} $$ \|
e^{-\tau{\cal L}} f \| \leq e^{-\tau \delta} \| f \|. $$ \end{enumerate}

The proof of Lemma 4.1 is based on \cite{7} (see also \cite{Wa}), and will
be given in the Appendix, where we also show that the scalar product
$(\psi_A,f)$ is well-defined for $f\in {\cal B}$.

\vs{3mm}\par\noindent Returning to the proof of Proposition 3, we write the
RG map, defined by (\ref{d5}), as \be a'_1 &=& L^{-2{\cal L}_A} a_1 + Ln_1
(Lx,L^2) \label{d9} \\ a'_2 &=& L^{-2{\cal L}} a_2 + L n_2 (Lx,L^2)
\label{d10} \en where \be n_1 (x,t) &=& -\int^{t}_1 ds \int dy G_A
(t,s,x,y)
(u_1 u^2_2 (y,s)-u_1A^2 s^{-1} \phi^2 (\frac{y}{\sqrt s}) ) \label{d11} \\
n_2 (x,t) &=& \int^{t}_1 ds \int dy G (t-s,x-y) u_1u^2_2 (y,s) \label{d12}
\en and $G_A$ is the fundamental solution of the $v_1$ equation in
(\ref{d6}) and $G$ is the kernel of $e^{d(t-s)\Delta}$, where we write
$\Delta$ for $\frac{d^2}{dx^2}$. Denote by $s_L$ the scaling $(s_L f) (x) =
L f(Lx)$ and by $G_A (t,s)$ the operator corresponding the kernel $G_A
(t,s,x,y)$. Then we have

\vs{5mm}\par\noindent {\bf Lemma 4.2} \begin{enumerate} \item[a)] $\| s_L
\|
\leq L$ \\ \item[b)]$\| G_A (t,s)\| \leq e^{c(t-s)}$;$\;\;\;\; \|
e^{d(t-s)\Delta} \| \leq e^{c(t-s)}$, $\;\;\;\;$ {\it for $c<\infty$}.
\end{enumerate}

\vs{5mm}\par\noindent {\bf Proof} $\;\;$ a) follows from $$ \sup_x L \;\;
|f(Lx)| (1+|x|)^q \;\; \leq L \|f\| $$ and b) from $$ 0 \leq G_A (t,s,x,y)
\leq G_0(t-s,x-y) $$ (which itself follows from the Feynman-Kac formula and
$A^2 \phi^2 \geq 0$), and the explicit Gaussian kernel of $G_0$. The kernel
of $e^{d(t-s)\Delta}$ is similar. \hfill$
\makebox[0mm]{\raisebox{0.5mm}[0mm][0mm]{\hspace*{5.6mm}$\sqcap$}}$ $
\sqcup$
\vs{3mm}

Let us now specify the $A$ in (\ref{d9}) and (\ref{d11}) (which is not the
same as the one in Theorem 1). We write \be a_1(x) &=& B \psi_A (x) + b_1
(x) \label{d13} \\ a_2 (x) &=& A \phi (x) + b_2 (x) \label{d14} \en with $$
B = (\psi_A, a_1),$$ $$A = \int a_2 dx. $$ Remembering that $\psi_A$ is
normalized by  $(\psi_A,\psi_A) = 1$, and $\phi$ by $\int \phi(x)dx =1$, we
see that $$ (\psi_A,b_1) = 0,$$ $$\int b_2 dx = 0. $$ Write $(a'_1,a'_2)$
similarily, with primes. The main estimate then is

\vs{5mm}\par\noindent {\bf Lemma 4.3} Given $L\geq L_0=e^{2\tau_0}$, with
$\tau_0$ as in Lemma 4.1,  there is an $\epsilon_0(L)>0$ such that if
$\|a_1\| \, \| a_2 \| < \epsilon \leq \epsilon_0 (L)$, we have
\begin{enumerate} \item[a)] $|A'-A| \leq C(L) \epsilon \| a_2 \|$ \item[b)]
$\| b'_2 \| \leq L^{-2\delta} \| b_2 \| +  C (L) \epsilon \| a_2 \|$
\item[c)] $| B' - L^{-2E_A} B | \leq C(L) \epsilon [\epsilon (1+ \| a_2 \|)
+ \| b_2 \| ]$ \item[d)] $\|b'_1\| \leq L^{-2(E_A + \delta)} \| b_1 \| +
C(L) \epsilon [\epsilon (1+ \| a_2 \|) + \| b_2 \| ]$ \end{enumerate} where
$C(L)$ is an $L$-dependent constant.

\vs{5mm}\par\noindent {\bf Proof} $\;\;$ We solve first $u_2$ from the
equation \be u_2(x,t) = e^{d(t-1) \Delta} a_2 + n_2 (x,t) \equiv u_{20} +
n_2 , \label{d15} \en with $n_2$ given by (\ref{d12}), by the contraction
mapping principle.  Consider the ball $$ B_R = \{u_2: | \| u_2 \| | \equiv
\sup_{t \in [1,L^2]} \| u_2 (\cdot,t) \| \leq R \| a_2 \| \}. $$ For $u_2
\in B_R$ we bound $n_2$ by using (\ref{b6}), i.e. $$ \| u_1 (\cdot,s)
\|_\infty \leq \| e^{d(s-1)\Delta} a_1 \|_\infty \leq \| a_1 \|_\infty \leq
C \| a_1 \| $$ and Lemma 4.2.b, to get \be |\| n_2 \| | \leq R^2 C(L) \|
a_1
\| \, \| a_2 \|^2 \leq R^2 C (L) \epsilon \| a_2 \|. \label{d16} \en Since
by Lemma 4.2.b again, $$ | \| u_{20} \| | \leq C(L) \| a_2 \|, $$ we see
that the right hand side of (\ref{d15}) maps $B_R$ into itself if $R =
R(L)$
is large enough and $\epsilon < \epsilon (L)$ is small enough. It is easy
to
see that the right hand side of (\ref{d15}) is a contraction in $B_R$, so
that we get a solution in $B_R$. By Lemma 4.2.a then, \be \| L n_2 (L
\cdot,L^2) \| \leq C(L) \epsilon \| a_2 \| \label{d17} \en and since \bea
A'
&=& A + \int L n_2 (Lx,L^2) dx \\ b'_2 &=& L^{-2 {\cal L}} b_2 + Ln_2
(Lx,L^2) + (A-A') \phi, \ena a) and b) follow from (\ref{d17}), and Lemma
4.1.e.

For $a'_1$, consider $n_1$ in (\ref{d11}), and write \be w(y,s) &\equiv &
u^2_2 (y,s) - A^2 s^{-1} \phi^2 (\frac{y}{\sqrt s}) \nonumber\\ &=& (u_2
(y,s) + A s^{-1/2} \phi (\frac{y}{\sqrt s})) ((e^{d(s-1) \Delta} b_2) (x) +
n_2 (x,s)), \label{d18} \en using (\ref{d15}), (\ref{d14}) and
$e^{d(s-1)\Delta} \phi (y)=s^{-\frac{1}{2}} \phi (\frac{y}{\sqrt{s}})$.
Thus, by (\ref{d16}) and Lemma 4.2.b, \be | \| w | \| \leq C(L) (\| a_2 \|
+
A) (\| b_2 \| + \epsilon \| a_2 \|) \label{d19} \en and, since $ A \leq C
\|
a_2 \|$, $s _L n_1 = Ln_1 (L \cdot, L^2)$ is bounded by \be \| s_L n_1 \|
\leq C(L) \epsilon (\| b_2 \| + \epsilon \| a_2 \|). \label{d20} \en (use
$\|u_1\|_\infty \leq C \| a_1 \|$ and $\|a_1\| \, \|a_2\| \leq
\epsilon)$.\\
Since, from (\ref{d9}) $$ B' = (\psi_{A'},a'_1) = (\psi_{A'} , L^{-{2\cal
L}_A} a_1) + (\psi_{A'}, s_Ln_1) $$ and, from (\ref{d13}) $$
(\psi_{A'},L^{-2{\cal L}_A} a_1) = B L^{-2E_A} (\psi_{A'},\psi_A) +
(\psi_{A'} , (P_{A'} -P_A)L^{-2{\cal L}_A} b_1) $$ (we used $P_A b_1 = 0$),
we get, using Lemma 4.1.c and (\ref{d20}), \be | B' - L^{-2E_A} B | &\leq &
C |A-A'| (B + \| b_1 \| ) + C(L)\epsilon (\| b_2 \| + \epsilon \| a_2 \|)
\nonumber \en where the constant $C$ in $C |A-A'|$ is independent of
$A,A'$,
because $A$ here is uniformly bounded (by Lemma 2.1). Then, using part a)
above, we get \be | B' - L^{-2E_A} B | &\leq & C(L) \epsilon [ (B + \| b_1
\| )\| a_2\| + \| b_2 \| + \epsilon \| a_2 \| ] \nonumber \\ &\leq & C(L)
\epsilon [ \epsilon (1+ \| a_2 \|) + \| b_2 \|)] \label{d21} \en (since
$B+\| b_1 \| \leq C \| a_1 \|$ and $\|a_1\| \, \|a_2\| \leq \epsilon$),
i.e.
we prove c).

Finally, for $b'_1$, write (use (\ref{d9}), (\ref{d13})) \bea b'_1 =
(1-P_{A'}) a'_1 &=& BL^{-2E_A} (P_A - P_{A'}) \psi_A + L^{-2{\cal L}_A} b_1
+ (P_A - P_{A'}) L^{-2{\cal L}_A} b_1 \\ & & + (1-P_{A'}) s_L n_1 \ena
(using again $P_A b_1 = 0$). Now, Lemma 4.1.b, c and (\ref{d20}) imply $$
\|
b'_1 \| \leq L^{-2(E_A + \delta)} \| b_1 \| + C(L) \epsilon [\epsilon
(1+\|a_2\|) + \| b_2 \|] $$ which is d). \hfill$
\makebox[0mm]{\raisebox{0.5mm}[0mm][0mm]{\hspace*{5.6mm}$\sqcap$}}$ $
\sqcup$ \vs{3mm}

For later purposes, we derive a lower bound for $B$. Recalling the
definition (4.18), write $$ u_{1t} = (\Delta-
{A'}^2t^{-1}\phi^2(\frac{x}{\sqrt t})) u_1 -  (w(x,t)
+(A^2-{A'}^2)t^{-1}\phi^2(\frac{x}{\sqrt t})) u_1 .$$ Using the Feynman-Kac
formula, we deduce $$ a'_1 (x) \geq (L^{-2{\cal L}_{A'}} a_1) (x)
e^{-C(L)(\||w\||+|A^2- {A'}^2|)} $$ and thus a lower bound \be B' \geq
L^{-2E_{A'}} B e^{-C(L)(\||w\||+|A^2- {A'}^2|)}. \label{d22} \en

\vs{5mm}\par\noindent {\bf Proof of Proposition 3}  We decompose $a^n_i$ as
in (\ref{d13}), (\ref{d14}) and derive bounds for $A_n,B_n$ and $b^n_i$
using Lemma 4.3. Set $$ nE_n = \sum^{n-1}_{m=0} E_{A_m}. $$ Note that $A_n
\geq A$, so that $E_{A_n} \geq E_A > 0$. Let $\eta < \min (\delta , E_A)$.
Then there exists a constant $C(L)$ (depending possibly on $L$ but not on
$n$) such that \be 0 \leq A_n &\leq & C(L) \| a_2 \| \non \\ \| b^n_2 \|
&\leq & C(L) L^{-2n\eta} \|a_2\| \non\\ 0 \leq B_n &\leq & C(L) L^{-2nE_n}
\| a_1 \| \non \\ \| b^n_1 \| &\leq & C(L) L^{-2n (E_n+\eta)} (\| a_1 \| +
\epsilon \|a_2\|) \non \\ \epsilon_n &=& \|a_1^n\| \|a_2^n\| \leq
C(L)L^{-2nE_n}\epsilon. \label{d23} \en The bounds (\ref{d23}) hold by
definition for $n=0$, and the induction follows from Lemma 4.3: the bound
on
$\epsilon_n$ follows from the first  four bounds in (\ref{d23}), and it
can,
in turn, be inserted in Lemma 4.3 to iterate those bounds. For $B_n$, we
iterate $B_n \leq C(L)(1-L^{-n\eta}) L^{-2nE_n} \| a_1 \|$ (which implies
(\ref{d23})), in order to control the right hand side in Lemma 4.3 c.
Furthermore, the bound on $\epsilon_n$ and Lemma 4.3 a imply that \be |
A_{n+1} - A_n | \leq C(L) L^{-2nE_n} \epsilon \|a_2\| \label{d24} \en and
thus $ A_n \to A^\ast$, for some $ A^\ast$; moreover, $$A^\ast = \int (a_1
+
a_2) dx, $$ because $\int (a_1 + a_2) dx$ is conserved (by Lemma 2.1) and
$\int a^n_1 dx \to 0$, by (\ref{d13}), (\ref{d23}).  Since $E_A$ is
continuous in $A$, by Lemma 4.1 c, \be E_{A_n} \to E^\ast = E_{A^\ast} ,
\;\; E_n \to E^\ast \label{d25} \en  From Lemma 4.3.c and (\ref{d23}), we
get that \be | B_{n+1} - L^{-2E_{An}} B_n | \leq C(L) \epsilon
L^{-2n(E_n+\eta)} [\epsilon + \| a_2 \|] \en and by (\ref{d25}), there
exists a $B^\ast$ such that \be B_n L^{2nE_n} \to B^\ast \label{d26} \en By
(\ref{d22}) and (\ref{d19}), (\ref{d23}), (\ref{d24}),
 $$ B_{n+1} \geq L^{-2E_{A_{n+1}}} B_n e^{-C(L) ( \|a_2 \|^2
L^{-2n\eta}+\epsilon\|a_2 \|e^{-2nE_n})} $$ so,  $B^\ast > 0$. Equations
(\ref{d23}), (\ref{d24}) and (\ref{d26}) may be rewritten, using
(\ref{d4}),
(\ref{d13}), (\ref{d14}), as:  \be \| \sqrt t u_2 (\sqrt t \cdot, t) -
A^\ast \phi \| \leq C t^{-\eta} \| a_2 \| \label{d27} \en \be \|
t^{\frac{1}{2} + E_{A^\ast}} u_1 (\sqrt t \cdot,t) - B^\ast \psi_{A^\ast}
\|
\leq C t^{-\eta} \epsilon (\epsilon + \| a_2 \|) \label{d28} \en  for times
$t=L^{2n},L>L_0$. For $t\in [L^{2n},L^{2n+2}]$ we use similar estimates for
the $n_i$ in (\ref{d11}) and (\ref{d12}), and, dropping the $\ast$, we get
(\ref{a6}, \ref{a7}). \hfill$
\makebox[0mm]{\raisebox{0.5mm}[0mm][0mm]{\hspace*{5.6mm}$\sqcap$}}$ $
\sqcup$

\vspace*{10mm}

\par\noindent {\large \bf Acknowledgements}

\vs{5mm}

We thank E. Titi for helpful discussions, and the Mittag-Leffler institute
for the hospitality that made the present collaboration possible. This work
was partially supported by NSF grants DMS-9205296 (A.K.), DMS-9302830
(J.X.), by EC grant CHRX-CT93-0411 (J.B. and A.K.), by Swedish NFR Grant
F-GF10448-301 (J.X.) and by the Finnish Academy (A.K.).

\vs{10mm}

\setcounter{section}{0} \appendix

\par\noindent {\Large \bf Appendix 1: Proof of Lemma 4.1.}

\vs{5mm}

\setcounter{equation}{0} First, observe that ${\cal L}_A$, acting on its
domain in $L^2 (R^1, d\mu)$,  is conjugated to a perturbation of the
Hamiltonian of the harmonic oscillator: $$ e^{\frac{x^2}{8}}{\cal L}_A
e^{-\frac{x^2}{8}}=  H_A \equiv -\frac{d^2}{dx^2}
+\frac{x^2}{16}-\frac{1}{4} +A^2\phi^2(x), \eqno{(A.1)} $$ acting in $L^2
(R^1, dx)$. Hence, ${\cal L}_A$
 has a compact resolvent, a pure point spectrum and, using the Feynman-Kac
formula and the Perron-Frobenius theorem \cite{GJ}, a non-degenerate lowest
eigenvalue. The same conclusions hold for ${\cal L}$.

To prove a), let us differentiate $$ {\cal L}_A \psi_A =E_A \psi_A
\eqno{(A.2)} $$ with respect to $A^2$. We get: $$ \phi^2 \psi_A +{\cal L}_A
\psi'_A = E'_A \psi_A +E_A\psi'_A. \eqno{(A.3)} $$ Now, we take the scalar
product of (A.3)  with $\psi_A$, and use $(\psi_A,\psi_A)=1$ (which implies
$(\psi_A,\psi_A')=0)$, to get $$ E'_A=(\psi_A, \phi^2 \psi_A). \eqno{(A.4)}
$$ Since $\phi>0$, $E'_A>0$, and, for $A=0$, we have
 $ \psi_0= \frac{e^{-\frac{x^2}{4}}}{\sqrt {4\pi}}$, and $E_0=0$.
Therefore,
$E_A>0$ for $A>0$.

To prove b), we discuss only the second claim, since  the first one is
similar but easier (and holds for any $q>1$). Observe that, since $E_A$ is
non-degenerate and $f$ is orthogonal to $\psi_A$, the bound would be
trivial
if we took the norms in ${\cal H}$. But functions in ${\cal H}$ have
essentially a Gaussian decay at infinity,  while those in ${\cal B}$ have a
polynomial decay. To go from a contraction in ${\cal H}$ to a contraction
in
${\cal B}$, we use an idea of \cite{7}. First notice that, since  $A^2
\phi^2 \geq 0$, the Feynman-Kac formula gives $$ e^{-\tau{\cal L}_A}
(x,y)\leq e^{-\tau{{\cal L}_0}}(x,y) \eqno{(A.5)} $$ and $e^{-\tau{{\cal
L}_0}}(x,y)$ is explicitly given by Mehler's formula \cite{Si}: $$
(e^{-\tau{\cal L}_0}) (x,y) = (4 \pi (1-e^{-\tau}))^{-\frac{1}{2}} \exp
\left(- \frac{(x-e^{-\tau/2}y)^2}{4(1-e^{-\tau})} \right) \eqno{(A.6)} $$
Hence, if a function $v$ satisfies $$ |v(x)| \leq C(1+|x|)^{-q},
\eqno{(A.7)}
$$ for some constant $C$, we have $$ |(e^{-\tau {\cal L}_0} v) (x)| \leq
 C'e^{\frac{\tau}{2}} (1+|x| e^{ \frac{\tau }{2}})^{-q} \eqno{(A.8)} $$ for
$|x|\geq 2 \sqrt {q \tau}$ and another constant $C'$. Hence, the operator
$e^{-\tau {\cal L}_0}$ contracts, for $|x|$ and $\tau$ large, any function
that decays as in (A.7) with $q>1$. By (A.5), we see that ${\cal L}_A$
behaves similarly.
 So, to prove b), we shall use the contraction in ${\cal H}$ for $x$ small
and (A.8) for $x$ large. However, we need here $q$ large, depending on
$E_A$,
 hence on $A$. For the other bounds in Lemma 4.1, any $q>1$ suffices.

Besides, let $\phi_n$ be the $n$th Hermite function which is an eigenvector
of $H_0$ in (A.1) (they are of the form $P_n (x) e^{-\frac{x^2}{8}}$,
where
$P_n$ is a polynomial of degree $n$).
 One can show that, for any $C>0$, for some even $n=n(A)$ and for any
$|x|$  large enough, $$ (H_A -E_A) ( C \phi_n - e^{\frac{x^2}{8}} \psi_A) >
0. $$ Indeed, $(H_A -E_A) e^{\frac{x^2}{8}} \psi_A =0$
 by (A.1, A.2) and  $H_A \phi_n \geq \frac{n}{2} \phi_n >0$ (because
$\phi_n
> 0$ for n even and $|x|$ large). Using the maximum principle, the
inequality $ \frac{x^2}{16} -\frac{1}{4}- E_A > 0$, for $|x|$ large, and
the
fact that there exists a large $|x|$ so that $C \phi_n - e^{\frac{x^2}{8}}
\psi_A > 0$,  for a sufficiently big $C$,
 one concludes that $\psi_A$ is bounded by:
 $$ 0\leq \psi_A (x) \leq C(A) (1+ |x|^n) e^{-\frac{x^2}{4}} \eqno{(A.9)}
$$
for some $n=n(A)$,  which implies that the scalar product $(\psi_A,f)$ for
$f\in {\cal B}$ is well-defined, if $q=q(A)$ is large enough.

To prove b), it is convenient to introduce the characteristic functions $$
\chi_s = \chi (|x| \leq \rho) $$ $$ \chi_\ell = \chi (|x|  > \rho) $$ where
$\rho$ will be chosen suitably below. The properties of  ${\cal L}_A$ that
we need are summarized in the following

\vs{3mm}

\no {\bf Lemma A.1} There exist constants $C<\infty, c>0$, such that
\begin{enumerate} \item[i)] For $g \in {\cal B}$, $$ \| e^{-\tau {\cal
L}_A}
g \| \leq C e^{\frac{\tau}{2}} \| g \|. \eqno{(A.10)} $$ \item[ii)] For $g
\in L^2 (R^1,d \mu)$, $$ \| e^{-{\cal L}_A} g \| \leq C \|g\|_2,
\eqno{(A.11)} $$ where $\| \cdot\|_2 $ is the norm in $L^2 (R^1,d \mu)$.
\item[iii)] For $g$ such that $\chi_s g \in L^2 (R^1,d \mu)$, $$ \|
\chi_\ell e^{-\rho {\cal L}_A} \chi_s g \| \leq  e^{- \frac{\rho^2}{5}} \|
\chi_s g \|_2, \eqno{(A.12)} $$ for $\rho$ large enough. \item[iv)] For $g
\in {\cal B}$, $$ \| \chi_\ell e^{-\rho {\cal L}_A} g \| \leq e^{- cq \rho}
\| g \|, \eqno{(A.13)} $$ for $\rho$ large enough, and $q>1$.
\end{enumerate}

\vs{3mm}

Now, take $f\in {\cal B}, (\psi_A,f)=0, \|f\| =1$.  We set $\tau_n=n\rho$,
and, using the Lemma, we  prove inductively that there exists a $\delta >
0$
such that $v(\tau_n)=e^{-\tau_n{\cal L}_A}f$ satisfies, for $\rho$ large,
$$
\| \chi_s v (\tau_n) \|_2 + \| \chi_s v(\tau_n)\| \leq e^{\frac{\rho^2}{6}}
e^{- \beta n}, \eqno{(A.14)} $$ and $$ \| \chi_\ell v (\tau_n) \| \leq e^{-
\beta n}. \eqno{(A.15)} $$ with $\beta=(E_A+\delta)\rho$. Part b) of Lemma
4.1. follows from (A.14), (A.15) by  taking a smaller $\delta$, in order to
bound the constants,
 for $\tau \geq \tau_0$, with $\tau_0$ large (for times not of the form
$\tau=n\rho$, use (A.10)). The bounds (A.14), (A.15) hold for $n=0$, for
$\rho$ large enough, using $\|f\| =1$ and the obvious inequality $$ \|
\chi_s f\|_2 \leq e^{\frac{\rho^2}{8}} \| f\|. \eqno{(A.16)} $$

So, let us assume (A.14), (A.15) for some $n \geq 0$ and prove it for
$n+1$.
Let $v = v(\tau_n)$ and write $$ v = \chi_s v + \chi_\ell v \equiv v_s +
v_\ell. $$  For all $n$, $(v(\tau_n),\psi_A)=0$, so that $|(v_s,\psi_A)|=
|(v_l,\psi_A)|\leq C(A)\rho^{-(q-1)}$, where we use (A.9) to
 derive the last inequality. Then, we get $$ \begin{array}{lll} && \|
e^{-\rho {\cal L}_A} v_s \|_2 \leq (C(A)\rho^{-(q-1)}
 e^{-\rho E_A} + e^{- \rho E'_A}) \| v_s \|_2 \nonumber\\ &\leq&
e^{-\rho(E_A+2\delta)} \|v_s\|_2\leq \frac{1}{4} e^{\frac{\rho^2}{6}}
e^{-\beta(n+1)} \end{array} \eqno{(A.17)} $$ where $E'_A$ is the second
lowest eigenvalue of  ${\cal L}_A$ and, in the second inequality, we choose
$\delta$ small, $q>\delta \rho$ and $\rho$ large. For the third inequality,
we used (A.14) and $\rho$ large, so that $e^{-\delta \rho}\leq \frac{1}{4}
$. Combining (A.11) and (A.17), we get $$ \| e^{-\rho {\cal L}_A} v_s \|
\leq C \| e^{-(\rho-1){\cal L}_A} v_s \|_2 \leq C e^{- (\rho-1) (E_A + 2
\delta)} e^{\frac{\rho^2}{6}} e^{-\beta n} \leq \frac{1}{4}
e^{\frac{\rho^2}{6}} e^{-\beta(n+1)} \eqno{(A.18)} $$ again, for
$e^{-\delta
\rho} $ small enough. Finally, from (A.10) and (A.15), we have $$ \|
e^{-\rho {\cal L}_A} v_\ell \| \leq C e^{\frac{\rho}{2}} e^{-\beta n}
\eqno{(A.19)} $$ and, from this and (A.16), we get $$ \| \chi_s e^{-\rho
{\cal L}_A} v_\ell \|_2 \leq C e^{\frac{\rho^2}{8}} e^{\frac{\rho}{2}} e^{-
\beta n}. \eqno{(A.20)} $$

Combining (A.17)-(A.20), one gets (A.14), with $n$ replaced by $n+1$ for
$\rho$ large  enough. On the other hand, (A.15), with $n$ replaced by
$n+1$,
follows immediately from (A.14), (A.15) and (A.12), (A.13), taking $cq >
E+\delta$. We choose the constants as follows: take $\delta$ small and
$\rho$ large and $q>\delta \rho > \frac{E+\delta}{c}$.

Turning to c), we observe that (A.4) and (\ref{a4}) imply that $E'_A \leq
(4\pi d)^{-1}$. Next, (A.3) implies that $$ \psi'_A= ({\cal L}_A -E_A)^{-1}
(E'_A-\phi^2)\psi_A \eqno{(A.21)} $$ where $$ ({\cal L}_A
-E_A)^{-1}=\int_0^\infty e^{-\tau({\cal L}_A -E_A)}d\tau \eqno{(A.22)} $$
is
a bounded operator on the subspace $\{f\in {\cal B}|  (\psi_A,f)=0\}$,
because of b) above. Also, (A.4) means that
 $(\psi_A,(E'_A-\phi^2)\psi_A)=0$. Hence, the norm of the  right hand side
of (A.21) is bounded, and we have $$ \|\psi_A-\psi'_A\|\leq C(A,A')
|A^2-A'^2|\leq C(K) |A-A'| \eqno{(A.23)} $$ for $A,A'\leq K$. Since (A.9)
shows that $P_A$ is well-defined and bounded in
 ${\cal B}$, point c) is proven. Point d) is an explicit computation (${\cal
L}\phi =0$),
 and the proof of e) is similar to the one of point b), since $\int fdx =
0$
means that $(f,\phi) = 0$ where $(\cdot,\cdot)$  is the scalar product in
$L^2 (R^1,e^{\frac{x^2}{4d}}dx)$ and $\phi$, given by (\ref{a4}), is the
principal eigenvalue of ${\cal L}$. \hfill$
\makebox[0mm]{\raisebox{0.5mm}[0mm][0mm]{\hspace*{5.6mm}$\sqcap$}}$ $
\sqcup$

\vs{3mm}

We are left with the

\vs{3mm}

\no {\bf Proof of Lemma A1}. Part (i) follows immediately from (A.5) and
(A.6).

For (ii), we use the Cauchy-Schwarz inequality applied to $$ (e^{-{\cal
L}_A}g) (x) = \int e^{-{\cal L}_A} (x,y)  e^{-\frac{y}{8}^2}
e^{\frac{y}{8}^2} g(y) dy $$ and the bound $$ \sup_x (1+|x|)^q \left(\int (
e^{-{\cal L}_A} (x , y)  )^2 e^{-\frac{y^2}{4}}d y \right)^{1/2} < \infty,
\eqno{(A.24)} $$ which follows from (A.5) and (A.6).

For (iii) proceed as in (ii) by using  Cauchy-Schwarz' inequality, but
replace (A.24) by $$ \sup_{|x| > \rho}(1+|x|)^q \left( \int | e^{-\rho{\cal
L}_A} (x,y)|^2 \chi (|y| \leq \rho) e^{-\frac{y}{4}^2} d y
\right)^{\frac{1}{2}} \leq e^{- \frac{\rho^2}{5}} \eqno{(A.25)} $$ which
again follows from (A.5) and (A.6)  (we can replace $\frac{1}{5}$ in (A.25)
by $\frac{1}{4} - \epsilon$ for any $\epsilon >0$, if $\rho$ is large
enough).

Finally, (iv) follows from (A.5) and (A.8).  (Since it is enough to have
$q>\delta \rho$ we have  $|x| > \rho \geq 2 \sqrt{q\rho}$ for $\delta$
small
and we can use (A.8)). \hfill$
\makebox[0mm]{\raisebox{0.5mm}[0mm][0mm]{\hspace*{5.6mm}$\sqcap$}}$ $
\sqcup$

\end{document}